\documentclass[english,preprintnumbers,amsmath,amssymb]{revtex4}
\usepackage[T1]{fontenc}
\usepackage[latin9]{inputenc}
\usepackage{float}
\usepackage{amsmath}
\usepackage{graphicx}

\makeatletter

\providecommand{\tabularnewline}{\\}

\@ifundefined{textcolor}{}
{%
 \definecolor{BLACK}{gray}{0}
 \definecolor{WHITE}{gray}{1}
 \definecolor{RED}{rgb}{1,0,0}
 \definecolor{GREEN}{rgb}{0,1,0}
 \definecolor{BLUE}{rgb}{0,0,1}
 \definecolor{CYAN}{cmyk}{1,0,0,0}
 \definecolor{MAGENTA}{cmyk}{0,1,0,0}
 \definecolor{YELLOW}{cmyk}{0,0,1,0}
 }

\usepackage{calc}

\makeatother

\usepackage{babel}

\begin{document}

\title{A new solution to the puzzle of the long lifetime of $^{14}$C}

\author{D. Robson}

\affiliation{Department of Physics, Florida State University, Tallahassee, 32306,
Florida, U.S.A.}
\begin{abstract}
A new cluster model solution to the long standing nuclear structure
problem of describing the anomalously long lifetime of $^{14}$C is
presented. Related beta-decay data for $^{14}\text{O}$ to states
in $^{14}\text{N}$, gamma decay data between low lying positive parity
states in $^{14}\text{N}$ and the elastic and inelastic magnetic
dipole electron scattering from $^{14}\text{N}$ data are all shown
to be very accurately described by the model. The shapes of the beta
spectra for the A=14 system are also well reproduced by the model.
The model invokes four-nucleon tetrahedral symmetric spatial correlations
arising from three- and four-nucleon interactions which yields a high
degree of $SU(4)$ singlet structure for the clusters and a tetrahedral
intrinsic shape for the doubly magic $^{16}\text{O}$ ground state.
The large quadrupole moment of the $^{14}\text{N}$ ground state is
obtained here for the first time and arises because of the almost
100\% d-wave deuteron-like-hole cluster structure inherent in the
model. 
\end{abstract}

\pacs{21.30.Fe, 21.60.Cs, 23.40.-s, 27.20.+n}

\maketitle

\section{INTRODUCTION}

The history surrounding the existing explanations of the $^{14}\text{C}$
lifetime has involved many attempts \cite{1,2,3,4,5,6,7,8,9,10},
but none of them have been completely satisfactory. As noted in the
most recent publications \cite{10} a full understanding of all the
data is expected to require three-nucleon interactions and/or clustering
considerations. Unfortunately, until now, the details of clustering
considerations and the nature of the three- (or more) nucleon interactions
have not been addressed. Much of the previous work has involved $A=14$
wave functions based on two $p$-shell holes in the closed shell reference
state (which is the $^{16}\text{O}$ ground state with $\{1s\}^{4}\{1p\}^{12}$
structure). This model is still the one used in the most recent structure
publication \cite{10}. Deviations from the simple two-hole shell
model have \cite{9} invoked additional multi-particle /multi-hole
states but the calculations show poor convergence and do not provide
any accurate description of the beta decay data. Other investigators
\cite{11,12} chose to use phenomenological admixtures of the possible
two-hole angular momentum configurations e.g., in \textit{L-S} coupling:
$\{{}^{3}\text{S}_{1},^{1}\text{P}_{1},^{3}\text{D}_{1}\}$ for the
$^{14}\text{N}$ $J^{P}=1^{+},T=0$ ground state and $\{{}^{1}\text{S}_{0},^{3}\text{P}_{0}\}$
for the $J^{P}=0^{+},T=1$ isospin states in $^{14}\text{C}$, $^{14}\text{N}$
and $^{14}\text{O}$. These type of attempts were criticized \cite{8,13}
for being inconsistent with the conventional strong j-j coupling shell
model. As we shall show the phenomenological approach appears to be
closer to the model used here and with more searching might have resulted
in the wave function admixtures presented here. 

In view of the failure of the shell model approach to provide a complete
description of the beta decay and allied data in the $A=14$ nuclei
it does appear that the shell model approach is \textit{not the } \textit{optimal
choice} as the basic picture for the $^{14}\text{C}$ beta decay problem.
That this appears to be the reality of the situation led this investigator
to invoke a more realistic model of the \textit{closed} \textit{shell}
nucleus $^{16}\text{O}$ which includes multi-nucleon correlations.
Such a correlated model has a longer history than the shell model
as it dates back to Wheeler \cite{14} in 1937. This was followed
by other investigators \cite{15,16} and for $^{16}\text{O}$ the
alpha-particle cluster model relying on the similarity with the methane
molecule $\text{CH}{}_{4}$ was used to describe the energy levels
as rotational vibrational states of a tetrahedral molecule in which
the H-atoms were replaced by alpha-particles and the C-atom at the
center was eliminated. All of the early work and almost all of the
later efforts with the alpha-particle model (see reference \cite{17}
for a review) have assumed the alpha clusters to be uncorrelated $\{1s\}^{4}$
configurations as in the simple spherical shell model. The lack of
correlations within each cluster and the assumption of spherical intrinsic
states leads to difficulties in obtaining accurate predictions with
the spherical alpha particle cluster model. In particular we know
of no attempts to describe the lifetimes of $^{14}\text{C}$ or $^{14}\text{O}$
using such models. 

Quite early in the history of clustering this investigator proposed
a model \cite{18} which synthesized the simple cluster model with
the shell model by introducing quark degrees of freedom into the bound
states of nuclei with $A=2,3$ and 4. The initial work (summarized
in \cite{19}) emphasized that many-body forces were to be expected
and that the three and four nucleon bound states would have spatial
correlations corresponding to point group symmetries $D_{3h}$ and
$T_{d}$ for $A=3$ and 4 respectively. A specific model \cite{20}
for the two-nucleon systems based on quark dynamics and one pion exchange
gave a realistic description of the deuteron and the phase-shifts
for low partial waves. The parameters of the model are consistent
with the one nucleon non-relativistic quark model and the basic symmetries
of QCD and chiral symmetry are adhered to. This approach was then
extended consistently \cite{21} to the spin independent part of the
three-nucleon interaction which showed that an \textit{equilateral
triangle} configuration with $D_{3h}$ point group symmetry for the
nucleons was strongly favored. The strong repulsive interactions of
up to 2 GeV between each pair of nucleons leads \cite{19} to a hole
in the charge density distribution at the nuclear center as originally
suggested by the authors of the experimental work \cite{22}. 

The $^{4}\text{He}$ ground state is expected to have \textit{tetrahedral
intrinsic} \textit{spatial }symmetry since it maximizes the three-
nucleon triangular configurations occurring on the four equivalent
faces of the tetrahedron. Indeed as indicated in \cite{19} the elastic
electron scattering data for $^{3}\text{He}$ and $^{4}\text{He}$
are very well described with equilibrium radii corresponding to triangular
and tetrahedral geometry. Again the hole at the center of the $^{4}\text{He}$
charge distribution is well described by the $T_{d}$ model. Such
spatial correlations are totally symmetric representations of the
orbital angular momentum rotation group $O(3)$ provided that the
intrinsic configuration is rotated through the three Euler angles
with equal weight and no parity change under inversion. These spatial
point group symmetries lead automatically to totally antisymmetric
$SU(4)$ states for spin $S$ and isospin $T$, i.e., $S=1/2=T$ for
$A=3$, and $S=0=T$ for the alpha particle. The $^{3}\text{He}$,
$^{3}\text{H}$ ground states are then simple one nucleon-hole states
in the alpha-particle. Here we focus on the alpha particle which has
$J^{P}=0^{+}$ and consequently $L=0$ only for its total orbital
angular momentum (as also occurs in the simple $\{1s\}^{4}$ shell
model configuration). At this point we note that the quark model discussed
above showed significant quenching of the one-pion exchange tensor
interaction as the nucleon- nucleon separation distance decreased
and for heavier meson exchanges between quarks there was essentially
no interaction. The $T_{d}$ spatial symmetry of the alpha- particle
intrinsic state in this model is therefore expected to lead to very
weak spin dependent contributions to the ground state cluster wave
function. In what follows we will assume only the leading $L=0=S=T$
state to be present in the ground state of $^{4}\text{He}$ and in
the intrinsic states of embedded four nucleon alpha-like clusters
in $^{16}\text{O}$.

\section{BASIS WAVE FUNCTIONS}

The cluster model wave functions used here show a strong resemblance
to the shell model basis states and indeed for $A=14$ the two-hole
states have the same total orbital $(L)$, spin $(S)$ and total angular
momentum J as those used in shell model states. For comparison purposes
we use the same notation as that of Genz \textit{et al} \cite{11}
and in \textit{L-S} coupling the most general wave functions are \begin{equation}
|^{14}\text{N},J^{P}=1^{+},T=0\rangle=\alpha\:{}^{3}S_{1}+\beta\:{}^{1}P_{1}+\gamma\:^{3}D_{1}\label{eq:1}\end{equation}
\begin{equation}
|^{14}\text{N}^{*},J^{P}=0^{+},T=1\rangle=\xi_{0}\:^{1}S_{0}+\eta_{0}\:^{3}P_{0}\label{eq:2}\end{equation}

\begin{equation}
|^{14}\text{C},J^{P}=0^{+},T=1\rangle=\xi_{-}\:^{1}S_{0}+\eta_{-}\:^{3}P_{0}\label{eq:3}\end{equation}
\begin{equation}
|^{14}\text{O,}J^{P}=0^{+},T=1\rangle=\xi_{+}\:^{1}S_{0}+\eta_{+}\:^{3}P_{0}\label{eq:4}\end{equation}
wherein only the $L^{P}=1^{+}$(denoted by $P$) states have the same
angular momentum substructure as the shell model uncorrelated $p^{-2}$
hole states. In what follows the angular momentum substructure of
the S and D cluster model states are not the same as the shell model
uncorrelated p$^{-2}$ hole states. The normalization of the above
four configurations however is the same, i.e., \begin{equation}
\alpha^{2}+\beta^{2}+\gamma^{2}=\xi_{\iota}^{2}+\eta_{\iota}^{2}=1\label{eq:5}\end{equation}
with $\iota=0,+\text{and }-.$ The values of $\xi_{\iota},\eta_{\iota}$
allow for a possible isospin triplet symmetry breaking which is needed
to describe the difference in the $\log(f_{A}t)$ values for the $\beta$$^{+}$
and $\beta$$^{-}$ decays of $^{14}\text{O}$ and $^{14}\text{C}$
leading to the ground state of $^{14}\text{N}$ respectively. 

The reference state for the cluster model hole states is also the
$J^{P}=0^{+},T=0$ ground state of $^{16}\text{O}$, but here we assume
the reference state is highly correlated with four alpha-like intrinsic
clusters with their centers of mass having equilibrium points at the
four corners of a tetrahedron as in Fig. 9 of \cite{19}. As in $^{4}\text{He}$
the ground state of $^{16}\text{O}$ is found by rotating over all
Euler angles with equal weight and requiring no change of parity under
inversion of all the spatial coordinates. The strong cluster correlations
between nucleons in the same alpha-like intrinsic cluster and the
much weaker correlations between two nucleons in different intrinsic
clusters leads to the assignment of $S$- and $D$-states for two
nucleons taken from the same cluster and P-states when the two nucleons
are taken from different clusters. In the cluster model presented
here the individual clusters have $SU(4)$ singlet structure and for
four clusters satisfying identical boson symmetry then the $^{16}\text{O}$
reference state will be pure \textit{L-S} coupled. Removing a pair
from an individual cluster allows the pair to have antisymmetric $SU(4)$
quantum numbers, which yields only $(S=0,T=1)$ and $(S=1,T=0)$ $SU(4)$
hole states. This in turn requires the relative motion of the two
nucleons to be in an even parity spherical harmonic $Y_{\ell}$ state.
As discussed earlier the $\ell$-value of the relative motion within
a tetrahedral cluster is taken here to be zero. The total orbital
angular momentum can be $L=0$ or 2 if the non-zero contribution comes
from the motion of the center of mass of the two nucleons relative
to the core. However when the two nucleons are taken from different
clusters the allowed $SU(4)$ pair states must be symmetric in order
to satisfy the bose permutation symmetry between the two identical
clusters. In this case the allowed $SU(4)$ states are $(S=0,T=0)$
and $(S=1,T=1)$ which requires their relative motion orbital angular
momentum to be odd valued. The ground states of the mirror nuclei
$^{15}\text{N}$ and $^{15}\text{O}$ have $J^{P}=1/2^{-}$ and as
in the shell model they are represented by proton and neutron $p_{1/2}$-hole
states in the ground state of $^{16}\text{O}$ respectively. In the
shell model the two p-holes can only have antisymmetric spatial states
if $L=1$ and this also holds true for the cluster model. Consequently
we assign the $P$ configurations to cluster model configurations
in which individual p-holes are taken in different clusters. For the
$T=1$ states in $A=14$ there are only two basis states: (a) the
$^{1}S_{0}$ state involving a pure $L=0$ dinucleon cluster extracted
from the same alpha-like cluster and (b) the $^{3}P_{0}$ state involving
a pure $L=1$ with each p-hole taken from a different alpha-like cluster.
Similarly for the $T=0$ basis states in $^{14}\text{N}$ there are
two types of basis states: (a) the $^{3}S_{1}$ and $^{3}D_{1}$ states
are pure $L=0$ and 2 deuteron- like clusters extracted from the same
alpha-like cluster and (b) the $^{1}P_{1}$ state with $L=1$ only
and with each p-hole coming from a different alpha cluster. 

The foregoing is important because it matters when considering the
energy matrix for the individual systems. It also matters when considering
observables that are dependent on the angular momentum substructure
of the orbital $L$-states or on their radial wave functions. This
is vital in the case of the quadrupole moment of the ground state
of $^{14}\text{N}$ and in describing elastic and inelastic electron
scattering data. The $\beta$-decay data, $M1$ gamma transitions
in $^{14}\text{N}$ and the magnetic moment of the ground state of
$^{14}\text{N}$ are independent of the substructures discussed above
as these observables depend only on the admixture amplitudes defined
in eqs. (\ref{eq:1}-\ref{eq:5}).

\section{ENERGY MATRICES }

For $T=1$ states with two basis states the $2\times2$ matrices involve
the coupling between the $^{1}S_{0}$ and $^{3}P_{0}$ states which
require a spin-orbit interaction which is antisymmetric in spin space
and also in orbital space. Such an interaction can be constructed
from the sum of one-body operators for each nucleon and is a superposition
of the nuclear spin-orbit $VN_{so}$ and the electromagnetic spin-orbit
$VE_{so}$ , see \cite{23}. Specifically these interactions for each
nucleon are\begin{equation}
VN_{so}\text{(neutron)}=\mbox{\ensuremath{\boldsymbol{\sigma}}}_{N}\cdot\text{\textbf{grad}}\{UN(\text{\textbf{r}}_{N})\}\times\text{\textbf{p}}_{N}=VN_{so}\text{(proton)}\label{eq:6}\end{equation}
\begin{equation}
VE_{so}\text{(neutron)}=[\mbox{\ensuremath{\boldsymbol{\sigma}}}_{N}\cdot\text{\textbf{grad}}\{UE(\text{\textbf{r}}{}_{N})\}\times\text{\textbf{p}}{}_{N}]\mu_{n}\label{eq:7}\end{equation}

\begin{equation}
VE_{so}\text{(proton)}=[\mbox{\ensuremath{\boldsymbol{\sigma}}}_{N}\cdot\text{\textbf{grad}}\{UE(\text{\textbf{r}}{}_{N})\}\times\text{\textbf{p}}{}_{N}](\mu_{p}-1/2).\label{eq:8}\end{equation}
 In these generalized spin-orbit interactions we assume that $UN(\text{\textbf{r}})$\textbf{
}and $UE(\text{\textbf{r}})$\textbf{ }are the nuclear and Coulomb
potentials which can involve tetrahedral harmonics with orbital-values
of 0, 3, 4, 6 etc. For convenience the nuclear terms in eq. \eqref{eq:6}
are taken to be the same for neutrons and protons and any differences
(which should exist ) are taken to be included in the overall magnitude
of the electromagnetic spin-orbit terms. The matrix elements for the
$T=1$ $2\times2$ matrix we denote by $H{}_{SS}^{1}$, $H_{PP}^{1}$
and $V_{SP}^{1}(=V_{PS}^{1})$, wherein the $S$ and $P$ subscripts
imply the$^{1}S_{0}$ and $^{3}P_{0}$ basis states. Values for these
matrix elements are different for $^{14}\text{C}$, $^{14}\text{N}$
$(T=1)$ and $^{14}\text{O}$ and for diagonal elements one has the
{}``unperturbed'' energies of the system whereas the off-diagonal
elements are simply the matrix elements of the spin-orbit interactions
given above. The $V_{SP}^{1}$ matrix elements are charge dependent
because the nucleon magnetic moments $\mu_{n}$ and $\mu_{p}$ are
of opposite sign. We characterize $V_{SP}^{1}$ for each member of
the isospin triplet for the nuclear spin-orbit matrix element by $v_{\text{nuc}}$
and the strength for the electromagnetic spin-orbit for two proton
holes by $v_{\text{el}}$. Consequently the values of the matrix elements
$V_{SP}^{1}$ for the isospin triplet are given by $v\text{\ensuremath{_{nuc}}}+f(N,N)v_{\text{el}}$
with $f(p,p)=1$, $f(n,n)=\mu_{n}/(\mu_{p}-1/2)$, $f(p,n)=(f(p,p)+f(n,n))/2$.
The amplitudes $\xi_{\iota},\eta_{\iota}$ for the three values of
$\iota=-,0\text{ and }+,$ corresponding to $^{14}\text{C}$, $^{14}\text{N}^{*}$
and $^{14}\text{O}$ respectively, are found by diagonalizing the
three $2\times2$ matrices when specific values of the unperturbed
energy differences $H_{PP}^{1}-H_{SS}^{1},$ $v\text{\ensuremath{\text{\ensuremath{_{nuc}}}}}$
and $v_{\text{el}}$ are used to fit the experimental data. In particular
the experimental values of the three energy differences between the
ground states of the isospin triplet and the corresponding first excited
states with $J^{P}=0^{+},T=1$ are constraints to be satisfied by
the eigenvalues of the three diagonalizations. These energy spacings
are 6.589 MeV in $^{14}\text{C}$, 6.305 MeV in $^{14}\text{N}^{*}$
and 5.91 MeV in $^{14}\text{O}$. This leaves two parameters out of
the five input parameters for the $T=1$ sector to be determined.

For the $J^{P}=1^{+},T=0$ case in $^{14}\text{N}$ there are three
basis states which can be labeled by $D,S\text{ and }P$ corresponding
to $^{3}D_{1},^{3}S_{1},$ and $^{1}P_{1}.$ We expect the $D$-state
to be the least bound state in the cluster model as it is even in
all shell models used historically. In the cluster model we expect
after diagonaliztion that the $D$-state occupation will be close
to 100\% and will be the ground state of $^{14}\text{N}$. The other
two eigenstates should be the $1^{+}$ states at 3.948 MeV and 6.204
MeV respectively. The diagonal matrix elements of the $3\times3$
energy matrix will have two unperturbed energy input values e.g.,
$H_{SS}^{0}-H_{DD}^{0}$ and $H_{PP}^{0}-H_{DD}^{0}$ and after diagonalization
the eigenvalue of the system should have the energy splittings of
3.948 MeV and 6.204 MeV respectively. The other three input values
are the matrix elements $V_{DP}^{0}$, $V_{SP}^{0}$ (which are non-zero
from the spin-orbit interaction in a similar manner to the $T=1$,
$V_{SP}^{1}$ discussed above) and $V_{DS}^{0}$ with the latter expected
to be very weak since it can only arise from a scalar product of rank
two tensors in spin and orbital spaces. This expectation is based
on the fact that tetrahedral symmetry does not have quadrupole harmonics
since this suppresses all tensor interactions in the $^{16}\text{O}$
system. The $T=0$ sector (like the $T=1$ sector) involves five input
parameters and fits to the energy differences reduces this number
of parameters to three which must be determined from other data. Thus
overall we have five variables in the energetics, two from the $T=1$
sector and three from the $T=0$ sector, and at first sight these
could be determined by the three $\beta$-decay data sets for $^{14}\text{C}(\beta^{-})$
and $^{14}\text{O}(\beta^{+})$ going to the ground and first excited
$J^{P}=1^{+}$ states in $^{14}\text{N}$. The five observables are
the three $\log(f_{A}t)$ values and the shapes of the $\beta$$^{-}$
and $\beta$$^{+}$ decays to the $^{14}\text{N}$ ground state. The
shape of the $\beta$$^{+}$ decay to the 3.948 MeV state in $^{14}$
N is not measured and in any event is expected to be constant as this
transition has a $\log(f_{A}t)$ of 3.138 corresponding to an unhindered
Gamow-Teller transition. Unfortunately, as shown by Towner and Hardy
\cite{24}, one needs to invoke renormalized axial $(g_{A})$ and
magnetic $(g_{l}$ and $g_{s})$ couplings to obtain the correct results
for the strongly hindered $\beta$- decay data and the radiative width
of the $T=1$ $^{14}\text{N}$ (2.313 MeV) state.

\section{RENORMALIZED OPERATORS}

The operators needed in the remainder of this paper are the free nucleon
coupling constants $g_{A},g_{lp},g_{ln},g_{sp}$ and $g_{sn}$ corresponding
to the axial vector $(g_{A}=1.2695)$, the orbital $g$-factors $(g_{lp}=1$
and $g_{ln}=0)$ and spin $g$-factors $(g_{sp}/2=2.79285$ and $g_{sn}/2=-1.91304)$,
with all the magnetic couplings being in units of nuclear magnetons
(n.m.). The renormalized $g_{A}$ for the $\beta$- decay studies
in $A=14$ is taken from the $^{15}\text{O}(\beta^{+})^{15}\text{N}$
mirror state transition as suggested by Towner and Hardy. The transition
is assumed to take place between single p-shell holes in the $^{16}\text{O}$
reference state and results in a renormalized value of $g_{A}^{*}=1.0885=(g_{A}-0.181)$
which yields the $\text{log}(\text{f}_{A}\text{t})=3.644$ for the
Gamow-Teller component \cite{25} of the $^{15}\text{O}(\beta^{+})^{15}\text{N}$
transition. The renormalized magnetic operators are taken here initially
to fit the magnetic moments of the ground states of the mirror nuclei
$^{15}\text{N}$ and $^{15}\text{O}$ based on a single p-orbital-hole
in the $^{16}\text{O}$ ground state. In looking at these magnetic
moments using the simple formulas for a p-hole with $j^{P}=1/2^{-}$
given by \begin{equation}
\mu_{i}=1/3(2g_{li}{}_{-}g_{si}/2)=g_{ji}/2\label{eq:9}\end{equation}
 where for $i=$free neutron $g{}_{nl}=0$ and for a free proton $g_{pl}=1$
and similarly $g_{ns}/2=-1.91304$ n.m. and $g_{ps}/2=2.79285$ n.m.
one finds values of the moments $g_{ji}/2$ as -0.26428 n.m. for $^{15}\text{N}$
and +0.63768 n.m. for $^{15}\text{O}$. These are not in good agreement
with the experimental values of -0.28319 n.m. for $^{15}\text{N}$
and +0.7189 n.m. for $^{15}\text{O}$. It is necessary to use renormalized
magnetic couplings as pointed out by Towner and Hardy \cite{24}.
They obtained these by including bound state shell model corrections
arising from core polarization and meson-exchange currents. 

In the cluster approach the p-hole arises by breaking the individual
alpha-like clusters in which it is embedded. This suggests that an
initial guess for the renormalized $g{}_{ji}^{*}/2$ values in the
A=15 systems should be given by \begin{equation}
g_{ji}^{*}(A=15)/2=g_{ji}(A=15)/2\times\{g_{ji}^{*}(A=3)/g_{ji}(A=3)\}\label{eq:10}\end{equation}
 in which $g_{ji}^{*}(A=3)/2$ are the observed moments for the s-hole
states in a free alpha particle which are -2.12750 n.m. and +2.97896
n.m. for the neutron and proton s-holes respectively. The values of
$g_{ji}(A=3)/2$ are taken to be the free nucleon magnetic moments.
The renormalized magnetic moments $\mu_{i}^{*}$ for $A=15$ are found
to be -0.28189 n.m. and +0.7092 n.m. for $^{15}\text{N}$ and $^{15}\text{O}$
respectively. This cluster model approach to the renormalization of
the magnetic moments of the nucleon-holes is remarkably accurate in
obtaining values of the $A=15$ magnetic moments which are within
0.5\% and 1.4\% of the observed values for $^{15}\text{N}$ and $^{15}\text{O}$
respectively. 

The above discussion for finding renormalized magnetic coupling constants
that describe the $A=15$ mirror states appears to be another validation
of the alpha-like cluster model for the $^{16}\text{O}$ reference
state. However further modifications to the values of $g_{li}^{*}$
and $g_{si}^{*}$ are needed to obtain a consistent and completely
accurate description of the magnetic moment data for the ground state
of $^{14}\text{N}$ as well as the A=15 mirror pair states. In using
the result from \eqref{eq:10} above we infer that $g_{lp}^{*}$ is
approximately 1.1 , $g_{ln}^{*}$ is approximately 0.0, $\mu_{sp}^{*}$
is approximately 3.0 and $\mu_{sn}^{*}$ is close to -2.15 (or if
$g_{ln}^{*}=0$ then $\mu_{sn}^{*}=-2.1567$ so that the magnetic
moment of $^{15}\text{O}$ is exactly reproduced). Best results are
achieved with the values: $g_{lp}^{*}=1.112$, $g_{ln}^{*}=0$, $\mu_{sp}^{*}=+3.0735664$
and $\mu_{sn}=-2.1567$ using the wavefunctions that fit the $\beta$
- decay data as discussed below.

Not only does this set of renormalized couplings fit the magnetic
moments of $^{14}\text{N}$ ($\mu=+0.403761\text{ n.m.}$), $^{15}\text{N}$
and $^{15}\text{O}$ but also exactly fits the magnetic moment of
the first $J^{P}=3^{-}$ state in $^{16}\text{O}$. The measured value
for this $3^{-}$ state is $\mu=+1.668$ n.m. corresponding to gJ
with the isoscalar gyromagnetic $g=+0.556$ (error is .005) . In the
tetrahedral model this $3^{-}$ state is a collective rotational excitation
of the $0^{+}$ ground state of $^{16}\text{O}$ for which the g factor
is $ $$Z/A=1/2$ if the orbital $g_{li}$ is 1 for protons and 0
for neutrons. Using the cluster renormalized orbital $g$- factors
given above yields $g=g_{lp}^{*}/2=+0.556$ in perfect agreement with
the data. In obtaining the magnetic moment for the ground state of
$^{14}\text{N}$ the shell model formula \cite{11}\begin{equation}
\mu=2^{-1/2}\{(\mu_{p}+\mu_{n})W_{1}/2^{-1/2}+W_{3}\}\label{eq:11}\end{equation}
(in which $W_{1}$, $W_{3}$ are $(2\alpha^{2}-\gamma^{2})$, $(\beta^{2}+3\gamma^{2}/2)/2^{-1/2}$
respectively), is modified for the cluster model to $\mu^{*}$ given
by \begin{equation}
\mu^{*}=2^{-1/2}\{\mu_{d}^{*}W_{1}/2^{-1/2}+g_{lp}^{*}W_{3}\}\label{eq:12}\end{equation}
 with $\mu_{d}^{*}$ being the renormalized isoscalar magnetic moment
of the deuteron-like hole state in $^{14}\text{N}$. The deuteron-hole
can involve internal orbital angular momentum of $l=0$ and 2 due
to the two-body tensor interaction between the neutron- and proton-holes
in the same alpha-like cluster. We use the equivalent parameterizations
for $\mu_{d}^{*}=f_{d}(\mu_{p}+\mu_{n})$ or $\mu_{d}^{*}=f_{d}^{*}(\mu_{p}^{*}+\mu_{n}^{*})$
the only new parameter is $f_{d}$ (or equivalently $f_{d}^{*}$)
since we use the renormalized nucleon-hole magnetic moments given
above. If $f_{d}^{*}$ is unity then the probability $P_{D}$ of any
$l=2$ state in the deuteron-hole is zero. For a small value of $P_{D}$
we should have $f_{d}^{*}$ being slightly less than unity. Using
the renormalized operators which fit the $A=15$ and $A=16$ magnetic
moments then the $^{14}\text{N}$ moment given by \eqref{eq:12} is
$0.382\text{ n.m.}$ if $f_{d}^{*}=1$ and is $0.403761\text{ n.m.}$
when we choose $f_{d}^{*}=.950706$ (or $f_{d}=.990755$) which is
consistent with a small value for $P_{D}$. The value of $P_{D}=8.3\%$
obtained here using\begin{equation}
P_{D}=(2/3)(1-f_{d}^{*})(\mu_{p}^{*}+\mu_{n}^{*})/(\mu_{p}^{*}+\mu_{n}^{*}-g_{lp}^{*}/2)\label{eq:13}\end{equation}
 is not inconsistent with any information concerning $D$-states in
the alpha particle. However by allowing for 1\% errors in the magnetic
moments of $^{15}\text{O}$ and $^{16}\text{O}$ one can obtain fits
to the magnetic moment of $^{14}\text{N}$ which have values of $P_{D}$
as low as 4.7\%.

\section{RESULTS FOR BETA-DECAY OBSERVABLES}

\begin{figure}
\includegraphics[width=3.4in]{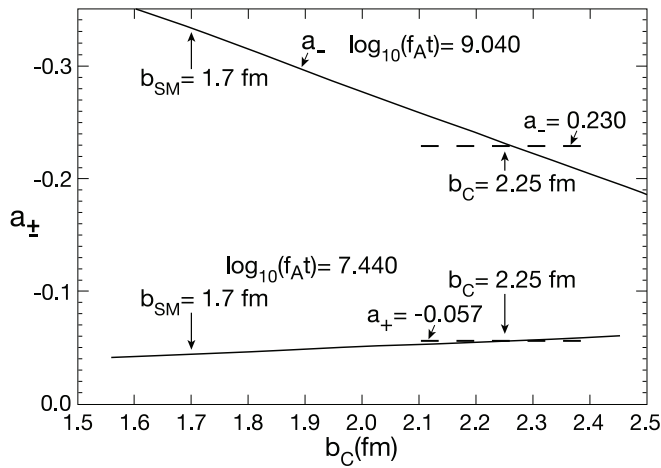}\caption{The slope parameters $a_{-}$ and $a_{+}$ used in \eqref{eq:16}
for $C(Z,W)$ for the $^{14}\text{C}(\beta^{-})$ and $^{14}\text{O}(\beta^{+})$
decays to the ground state of $^{14}\text{N}$ respectively as a function
of the average oscillator length $b_{c}.$ The $\log(f_{A}t)$ values
are independent of $b_{c}$ and correspond to the tabulated wavefunction
admixtures given in section \textbf{V.}}
\label{fig:1}

\end{figure}

For Gamow-Teller (GT) transitions in the $A=14$ nuclei we restrict
our considerations to $J^{P}=0^{+}$ initial states leading to $J^{P}=1^{+}$
states in $^{14}\text{N}$. The GT matrix element is given by \cite{24}
\begin{equation}
\text{MGT}=g_{A}^{*}6^{1/2}(\xi_{\iota}\alpha-\eta_{\iota}\beta/3^{1/2})\label{eq:14}\end{equation}
 where we include the renormalized axial coupling $g_{A}^{*}$ in
the definition of MGT. With this definition the $f_{A}t$ is given
by \cite{24} \begin{equation}
f_{A}t=6146/|\text{MGT}|^{2}\text{s}.\label{eq:15}\end{equation}
 in which the $f_{A}$ rate functions are corrected to include the
effects of the nuclear structures via the shape function $C(Z,W)$
used by Towner and Hardy \begin{equation}
C(Z,W)=|\text{MGT}|^{2}k(1+aW+\mu_{1}\gamma_{1}b/W+cW^{2})\label{eq:16}\end{equation}
 following the format of Genz \textit{et al }that involves the $a,b,c$
variables which are dependent on the details of the nuclear structure
model. To the accuracy needed for $A=14$ beta-decay shapes we use
$\mu_{1}\gamma_{1}=1$. The parameter $k$ is used to fit the data
and $W$ is the total electron (positron) energy in units of the electron
rest mass energy. The data \cite{27} for the $^{14}\text{C}(\beta^{-})^{14}\text{N}$
decay gives information on the slope parameter ($a$) whereas the
data \cite{28} for the $^{14}\text{O}(\beta^{+})^{14}\text{N}$ decay
gives information on all three parameters ($a,b$ and $c$) . The
expressions for $a,b$ and $c$ as well as the nuclear matrix elements
are given in detail in Genz et al \cite{11} and also in Garcia and
Brown \cite{26} who noted that the term denoted by $V_{4}$ has the
opposite sign from that given by Genz et al. Our calculations use
the sign choice of Garcia and Brown for $V_{4}$ as it is consistent
with the relation between 2-hole states and 2-particle states. We
choose not to include these lengthy relationships as they are readily
available in \cite{11,26}. It is important to note that the shape
functions calculated using the formulas in Genz \textit{et al }have
some terms which are model dependent. In particular we have used renormalized
$g^{*}$ for axial and magnetic couplings in place of the free nucleon
$g\text{'s}$ and also replaced the oscillator length ($b=1.7\text{fm}$
in Genz \textit{et al}) by the cluster model $b_{c}$ value obtained
from the cluster model fit (using linear combinations of p-state radial
harmonic oscillator states) to the inelastic electron scattering data
as discussed in VI below. The shape function is sensitive to the choice
of $b_{c}$. The behavior of the slope parameters $a_{-}$ and $a_{+}$
for $^{14}\text{C}(\beta^{-})$ and $^{14}\text{O}(\beta^{+})$ decaying
to the $^{14}\text{N}$ ground state using the wavefunctions shown
below are shown in Fig. \ref{fig:1} as a function of the average
oscillator parameter which is denoted by $b_{c}$ in Fermi units.

The cluster model wavefunctions that fit all the data in terms of
the coefficients $\alpha$, $\beta$, $\gamma$, $\xi_{\iota},$ $\eta_{\iota}$
are given for $T=0,J^{P}=1^{+}$and for $T=1,J^{P}=1^{+}$ in Table
\ref{tab:1} and Table \ref{tab:2} respectively.

\begin{table}[H]
\begin{tabular}{cccc}
 & $\alpha$ & $\beta$ &  $\gamma$\tabularnewline
\hline
$|^{14}\text{N},E=0\rangle$ & .0169003 & .1860092 & .9824026\tabularnewline
$|^{14}\text{N},E=3.948\rangle$ & .7600690 & -.6407632 & .1082473\tabularnewline
$|^{14}\text{N},E=6.204\rangle$ & .6496225 & +.7448645 & -.1522089\tabularnewline
\hline
\end{tabular} 

\caption{Numerical values of $\alpha,\beta,\gamma$ for the three $1^{+}$
states in $^{14}\text{N}$ which result from the diagonalization of
the symmetric $3\times3$ matrix as discussed in the text.}
\label{tab:1}
\end{table}

\begin{table}[H]
\begin{tabular}{ccc}
 & $\xi_{\iota}$ & $\eta_{\iota}$\tabularnewline
\hline
$|^{14}\text{C},E=0\rangle$ & .9865400 & .1635201\tabularnewline
$|^{14}\text{N},E=2.313\rangle$ & .9909549 & .1341954\tabularnewline
$|^{14}\text{O},E=0\rangle$ & .9945400 & .1043556\tabularnewline
$|^{14}\text{C},E=6.589\rangle$ & -.1635201 & .9865400\tabularnewline
$|^{14}\text{N},E=8.616\rangle$ & -.1341954 & .9909549\tabularnewline
$|^{14}\text{O},E=5.910\rangle$ & -.1043556 & .9945400\tabularnewline
\hline
\end{tabular} 

\caption{Numerical values of $\xi_{\iota},\eta_{\iota}$ for the two $0^{+}$
states in each of the isospin triplet of $A=14$ nuclei which result
from the diagonalization of the symmetric $2\times2$ matrix as discussed
in the text.}
\label{tab:2}
\end{table}

These structure coefficients are the eigenstates obtained by diagonalizing
the symmetric $3\times3$ matrix for $T=0$ with off-diagonal matrix
elements (in MeV) :$V_{SD}^{0}=-0.28864$, $V_{PD}^{0}=-0.9772$ ,
$V_{SP}^{0}=1.079353$ and diagonal unperturbed energies $E_{SS}=4.7088$,
$E_{PP}=4.873$ and $E_{DD}=0$. Similarly there are three symmetric
$2\times2$ matrices for the $T=1$ systems and their matrices have
elements given by Table \ref{tab:3} in which the unperturbed energies
have an average value for the isospin triplet which is 6.03 MeV, that
is remarkably similar to the energy splitting of the first two $J^{P}=0^{+}$
states in the reference system of $^{16}\text{O}$. 

\begin{table}[H]
\begin{tabular}{cccc}
 & $V_{SP}^{1}$ & $E_{SS}$ & $E_{PP}$\tabularnewline
\hline
$^{14}\text{C}$ &  -1.062993 & 0 & 6.237\tabularnewline
$^{14}\text{N}$ & -.8381848 & 0 & 6.076\tabularnewline
$^{14}\text{O}$ & -.6133766 & 0 & 5.7813\tabularnewline
\hline
\end{tabular}\caption{Numerical values of the elements of the $2\times2$ energy matrix
for the isospin triplet of $A=14$ nuclei.}
\label{tab:3}
\end{table}

The major effect of the electromagnetic spin-orbit in the $T=1$ isospin
triplet states lies in the relatively large differences of the $\eta_{\iota}$
coefficients. In particular the magnitude of $\eta$$_{+}$ for the
$^{14}\text{O}$ ground state is about 36\% smaller than $\eta_{-}$
for the mirror state in $^{14}\text{C}$. Indeed it is this large
difference in the cluster model $\eta$ coefficients which leads to
an explanation of the large ratio of the MGT elements for $^{14}\text{O}(\beta^{+})/^{14}\text{C}(\beta^{-})$.
It is vital to understand that it is the cluster correlations between
the nucleons that damps the nuclear spin-orbit $(v_{\text{nuc}})$
and enhances the electromagnetic spin-orbit $(v_{\text{el}})$. The
values used here are $v_{\text{nuc}}=-0.817874\text{ MeV}$ and $v_{\text{el}}=-0.245119\text{ MeV}$.
As defined above $v_{\text{el}}$ is appropriate for the two proton-hole
$^{14}\text{C}$ system and the values of $V_{SP}^{1}$ are given
using the superpositions of $v_{\text{nuc}}+f(N,N)v_{\text{el}}$
as given in \textbf{III }above. Although the symmetry breaking in
$\eta_{+}$ and $\eta_{-}$ is large the overlap of the $^{14}\text{C}$
and $^{14}\text{O}$ ground states is 0.99822 which involves an overall
symmetry breaking of less than 0.2\%. 

\begin{figure}

\includegraphics[width=3.4in]{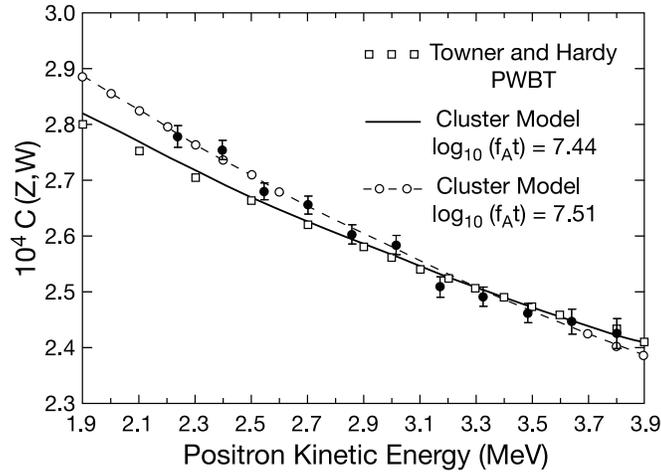}\caption{The shape function $C(Z,W)$ for the $^{14}\text{O}(\beta^{+})^{14}\text{N}(E=0)$
transition as a function of the positron kinetic energy (MeV). The
data points with error bars are taken from \cite{28} including corrections
given in \cite{24}.}
\label{fig:2}
\end{figure}

\begin{figure}

\includegraphics[width=3.4in]{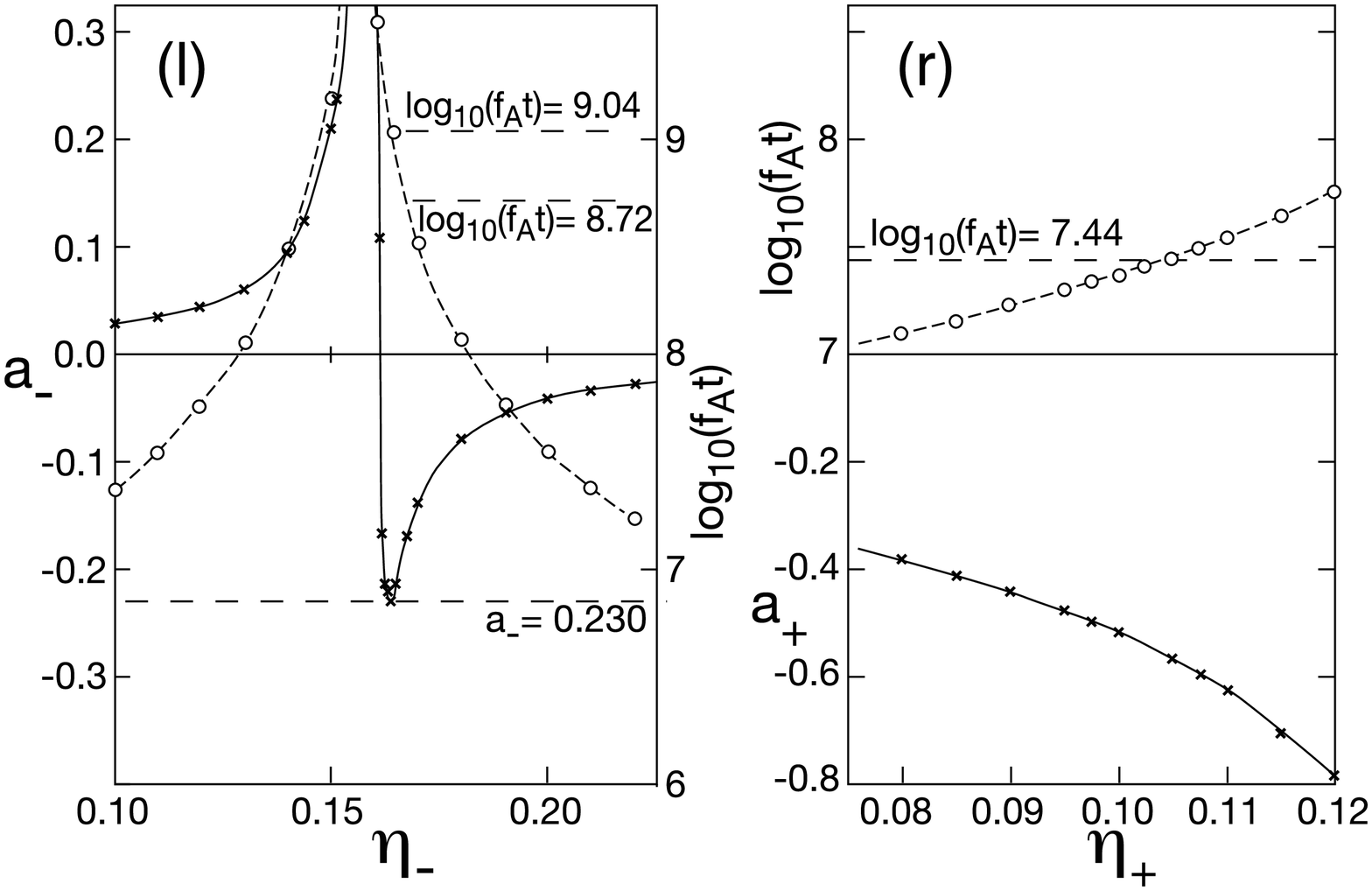}\caption{The values of $a_{-}$ and $a_{+}$ as a function of the respective
initial state admixture coefficients denoted by $\eta$$_{-}$ and
$\eta_{+}$ are shown by the continuous lines in the left and right
side panels respectively. The corresponding values of the $\log(f_{A}t)$
are shown by the dashed lines and for the $^{14}\text{C}(\beta^{-})$
case shows the strong variation in $a_{-}$ and $\log(f_{A}t)$ for
small changes in $\eta_{-}$ near $0.1635201.$}
\label{fig:3}

\end{figure}

Using the above wave functions and $b_{c}=2.25\text{ fm}$ leads to
the values of MGT, $\log(f_{A}t)$ and the shape parameters $a,b,c$
given in Table \ref{tab:4}. %
\begin{table}[H]
\begin{tabular}{cllllll}
\hline 
Model & MGT & $\log(f_{A}t$) & $k$ & $a$ & $b$ & $c$\tabularnewline
\hline 
$^{14}\text{O}(\beta^{+})^{14}\text{N}(E=0)$ &  &  &  &  &  & \tabularnewline
Cluster  & 0.01493 &  7.440 & 1.600 & -0.056 & 0.033 & 0.002\tabularnewline
Cluster  & 0.01373 & 7.513 & 1.990 & -0.060 & 0.036 & 0.003\tabularnewline
PBWT  &  0.01480 & 7.448 & 1.605 & -0.054 & 0.035 & 0.002\tabularnewline
Expt:$^{14}\text{O}$ & 0.018(0) & 7.284(7) &  &  &  & \tabularnewline
\hline 
$^{14}\text{O}(\beta^{+})^{14}\text{N}(E=3.948)$ &  &  &  &  &  & \tabularnewline
Cluster & 2.119 & 3.136 & 1.0 & 0.0  & 0.0  & 0.0 \tabularnewline
Expt:$^{14}\text{O}$ & 2.119(39)  & 3.138(16) &  &  &  & \tabularnewline
\hline 
$^{14}\text{C}(\beta^{-})^{14}\text{N}(E=0)$ &  &  &  &  &  & \tabularnewline
Cluster  & -0.00237 & 9.040 & - & -0.231 & 0.125 & 0.023\tabularnewline
Cluster  & -0.00343 & 8.718 & - & -0.176 & 0.063 & -0.003\tabularnewline
PBWT  & -0.00343 & 8.718 & 0.607 & -0.235 & 0.013 & 0.013\tabularnewline
Expt:$^{14}\text{C}$ & 0.002(0) & 9.040(3)  &  & \multicolumn{3}{c}{-0.23(2)  (100-160 keV)}\tabularnewline
 &  &  &  & \multicolumn{3}{l}{-0.17 (50-160 keV) }\tabularnewline
\hline
\end{tabular}

\caption{$\beta^{+}$ and $\beta^{-}$ transition results for MGT, $\log(f_{A}t$)
and $k,a,b,c$ shape parameter values using cluster wave functions
from the text and PWBT shell model results (with their renormalized
operators) from Towner and Hardy \cite{24}. Expt. values for $|\text{MGT}|$
and $\log(f_{A}t$) are from \cite{25}. The slope parameter a for
$^{14}\text{C}$ is taken from Table II of \cite{27} using the fitted
value of $a=-0.45(4)\text{ MeV}^{-1}$ in the energy range $100-160\text{ keV}$.
In electron rest mass units this becomes $a=-0.23(2)$. For the wider
energy range a in electron rest mass units is $-0.17$.}
\label{tab:4}

\end{table}
For the special case of $^{14}\text{O}(\beta^{+})^{14}\text{N}(E=0)$
the present work yields almost identical results to those of Towner
and Hardy (who are the only ones to use renormalized couplings). The
fit to the experimental data they obtained for the shape function
using the PWBT \cite{29} shell model basis shown in Fig. \ref{fig:2}
has a slightly larger $\chi^{2}$ than that obtained with the cluster
model. %
The best fit to the data is obtained using a small change in $\eta_{+}$
to 0.1085 which gives the dashed curve in Fig. \ref{fig:2} and corresponds
to a $\log(f_{A}t)$ of 7.51 and an MGT of 0.01373 for the $^{14}O(\beta^{+})$
transition.The sensitivity of the slopes $a_{+}$ and $a{}_{-}$ and
the corresponding $\log(f_{A}t)$ values to the values of $\eta{}_{+}$
and $\eta_{-}$ is shown in Fig. \ref{fig:3}. The sensitivity of
$a_{+}$ to $\eta_{+}$ is very much less than the sensitivity of
$a_{-}$ to $\eta_{-}$ which in the cluster model is very sensitive.

The cluster $^{14}\text{C}(\beta^{-})^{14}\text{N}(E=0)$ results
are significantly different from those of Towner and Hardy as our
wavefunctions give the usual value of 9.04 for the $\log(f_{A}t)$
when the slope parameter $a_{-}=-0.231$ rather than the PWBT shell
model $\log(f_{A}t)$ of 8.72. We have been unable to exactly pin
down the source of this difference. We note that the main thrust of
the Towner and Hardy \cite{24} investigation was to show that CVC
could be satisfied accurately by using renormalized $g$ factors and
in the case of $^{14}\text{O}(\beta^{+})^{14}\text{N}_{gs}$ by adjusting
the PWBT ground state of $^{14}\text{N}$ they obtained a good description
of the $\beta^{+}$ shape function. We are concerned however \textit{that
the Towner and Hardy } \textit{calculations} \textit{used a different
wavefunction for $^{14}\text{N}(E=0)$ in the $^{14}\text{C}$ decay
from the one used in the} \textit{$^{14}\text{O}$ decay} \textit{and
recently Negret et al }\cite{30} \textit{implied that Towner and
Hardy had been able to account } \textit{for the} \textit{large asymmetry
in the mirror log(ft) values. Physically these mirror decays have
the same} \textit{final state in $^{14}\text{N}$ and one cannot explain
the large ratio of their MGT values by simply modifying } \textit{the
final state}.\textit{ Indeed Towner and Hardy did not claim} \cite{24}\textit{
to have explained the large asymmetry} \textit{in these mirror $\log(ft)$
values; their focus was on reconciling the $^{14}\text{O}$ shape-correction
form factor } \textit{with the M1 matrix element in $^{14}\text{N}$.
As pointed out by early workers }\cite{2,6}\textit{ the symmetry
breaking of these mirror transitions must arise from symmetry breaking
interactions in the initial mirror states. Here we have specified
that it is primarily the electromagnetic spin-orbit interaction in
the} \textit{initial states that causes the large asymmetry.} It is
also worth noting the opposite signs of the MGT values for these mirror
transitions because the shape parameters do not come out correctly
unless the mirror MGT values\textit{ }have opposite signs. This is
most readily seen in Fig. \ref{fig:3} where the values\textit{ }of
$\eta_{-}$ between 0.14 and 0.15 can yield $\log(f_{A}t)$ values
between 8.5 and 9.4 with MGT > 0 but also\textit{ } yield a positive
slope\textit{ }for $a_{-}$ in complete contradiction to experimental
data. 

One of the difficulties remaining is the uncertainties in the experimental
data \cite{27} for the $^{14}\text{C}(\beta^{-})$ transition. In
particular the value of the slope parameter depends strongly on the
range of electron energies used as indicated in Table \ref{tab:4}.
Also the accuracy of the data apparently did not allow any information
to be determined for the $b$ or $c$ coefficients in $C(Z,W)$ so
that a linear form $C_{L}(Z,W)=(1+a^{*}W)$ was used \cite{27} to
extract the effective slope $a^{*}$ for each energy range. The value
of $a^{*}=-0.45(4)\text{ MeV}{}^{-1}$ for the 100-160 keV range is
apparently believed to be the favored value for $a^{*}$ in ref. \cite{27}.
In the case of the shell model calculations given in ref. \cite{24}
one can least squares fit the models shape functions over the energy
range 100-160 keV with the linear form to obtain the effective $a^{*}$.
We find $a^{*}=-0.203,-0.214,$ and $-0.195$ for the three models
labeled CK, PWBT, and MK respectively in Table III of Towner and Hardy.
The corresponding values of a are $-0.215,$$-.235$ and $-.207$
which are only $6\%-9\%$ higher than their respective $a^{*}$ values.
For the cluster model with $\log(f_{A}t)=9.040$ and $a=-0.231$ a
linear fit to $C(Z,W)$ gives $a^{*}=-0.232$ because of the strong
cancellation of the $b$ and $c$ terms in this case. The cluster
case where the $\log(f_{A}t)=8.718$ and $a=-0.176$ when linearized
yields $a^{*}=-0.207$ which is consistent with the shell model values
for $a^{*}$. If one accepts the best value for $a^{*}=-0.23$ then
the cluster wavefunction with $\log(f_{A}t)=9.040$ is favored over
all the other models. Apparently only more accurate data for the $^{14}\text{C}(\beta^{-})$
transition can provide the necessary information on $a,$$b$ and
$c,$ or $a^{*}$. 

The $\beta^{+}$ decay of $^{14}\text{O}$ leading to the $^{14}\text{N}$
$(E=3.948)$ state calculated here has a $\log(f_{A}t)$ of 3.14 and
an MGT of 2.119 in perfect agreement \cite{25} with experiment. This
is not discussed by anyone else except for Genz \textit{et al} \cite{11}
who quoted that their model yielded a $\log(f_{A}t)$ value of 2.87.
In our case this latter transition was part of the fitting procedure
whereas Genz \textit{et al} did not include this transition as part
of their fitting procedure. However we believe that calculations of
this transition to the first excited $1^{+}$ state in $^{14}\text{N}$
are an additional restraint that all models should include. 

The major reason that renormalized couplings were used is to understand
whether the radiative $(M1)$ width of the $^{14}\text{N},T=1$ state
at $E=2.313$ is consistent with the model wave functions. Again the
model states above yield a value of $\Gamma_{\gamma}=6.7\text{meV}$
but only if renormalized values of the isovector coupling constants
from above are used thereby satisfying the conserved vector current
requirements. The formulas are given by Garcia and Brown \cite{26}
in their eq.(25) in terms of the structure coefficients $V_{1}$ and
$V{}_{3}$ given in \cite{11} and with our renormalized magnetic
couplings one needs in order to obtain $\Gamma_{\gamma}$ that 

\begin{equation}
|2^{-1}{}^{/2}(\mu_{p}^{*}-\mu_{n}^{*})V_{1}(\gamma)+g_{lp}^{*}V_{3}|=0.256\text{ n.m.}\label{eq:17-1}\end{equation}
 with an associated error of 0.006 n.m. The value obtained here is
0.256 n.m. since the radiative width $\Gamma_{\gamma}$ of 6.7(3)
meV was part of our fitting procedure for the coefficients $\alpha$,
$\beta$, $\gamma$, $\xi$$_{0}$ and $\eta$$_{0.}$ The recent
work of Holt \textit{et al }\cite{10} also considered the recent
experiments \cite{30} that determined $\text{BGT}=|\text{MGT}|^{2}/(2J_{i}+1)$
values from the $^{14}\text{N}$ ground state to excited states of
$^{14}\text{C}$ and $^{14}\text{O}$ using the charge exchange reactions
$^{14}\text{N}(\text{d}^{2},\text{He})^{14}\text{C}$ and $^{14}\text{N}(^{3}\text{He,t})^{14}\text{O}$.
For the three final states in each of $^{14}\text{C}$ and $^{14}\text{O}$
labeled as $0_{1}^{+},$$0_{2}^{+}$ we have wavefunctions as listed
above and for the $1_{1}^{+}$ states at 11.31 MeV and 11.24 MeV respectively
we use a pure $^{3}P_{1}$ two nucleon-hole in $^{16}\text{O}$ as
did Amos \textit{et al }\cite{13} . For the $0_{1}^{+}$ states the
beta decay data is more reliable and we already fitted their BGT values.
For the $0_{2}^{+}$ states we obtain $\text{BGT}=0.028$ for both
these transitions which are similar to the very small values shown
in Negret \textit{et al} \cite{30}. By using the definition in Holt
\textit{et al }for B(GT) corresponding to the inverse transition (note
that $g_{A}$ is not included in their definition of B(GT)) we obtain
for the $^{14}\text{C}$ case that $\text{B(GT)}=.071$ which appears
to agree closely with the experimental value shown in their Fig. \ref{fig:2}
for this transition. The Holt \textit{et al} theory\textit{ }result
for this transition is at least three times too large even with the
modified tensor interaction included. The $M1$ transition from the
ground state of $^{14}\text{C}$ to the $1_{1}^{+}$ state at $11.3\text{ MeV}$
in $^{14}\text{C}$ was first discussed in \cite{13} as a way to
obtain information on the $^{3}P_{0}$ component in the ground state
of $^{14}\text{C}$. The radiative width $\Gamma_{\gamma}$ of this
$1_{1}^{+}$ state was measured \cite{31} to be 6.8 meV by extrapolating
the inelastic electron scattering from $^{14}\text{C}$ data to the
photon point However such extrapolations can be quite inaccurate and
it is more reliable for this $1_{1}^{+}$ state to consider the BGT
measurements of Negret \textit{et al}. The calculated value of BGT
to this state in the cluster model is $\text{BGT}=0.082$ or $\text{B(GT)}=0.069$
which appears to be only about $30\%$ below the experimental value. 

The situation for the transitions to $J^{P}=2^{+}$ states from the
$^{14}\text{N}$ ground state is beyond the scope of this work as
it is too complicated for us to calculate with any accuracy in the
cluster model since there are three $2^{+}$ states observed in each
mirror system with significant strength and only two $2^{+}$ states
with the simple two-hole structure in the $^{16}\text{O}$ ground
state. However the two simple states , $^{1}D_{2}$ and $^{3}P_{2},$
will have a strong BGT only for the $^{1}D_{2}$ because the initial
state has over $98\%$ in amplitude in the $^{3}D_{1}.$ This means
that the transitions to the three mixed $2^{+}$ states observed will
dominate the GT transitions as suggested by Aroua \textit{et al }\cite{9}
and observed in the experiments of Negret \textit{et al}.

\section{ELASTIC AND INELASTIC M1 ELECTRON SCATTERING}

In this section it is necessary to go beyond the admixture coefficients
 $\alpha,\beta,\gamma,\xi_{\iota},\eta_{\iota}$ and the angular momentum
quantum numbers because the form factors for $M1$ transitions involve
momentum transfer $(q)$ dependence and four types of multipoles for
each nucleon \cite{32}. In the $p^{-2}$ simple shell model it is
customary to use a single p-shell harmonic oscillator wave function
for each nucleon and this yields very simple spherical Bessel transforms
for $L=0$ and $L=2$ amplitudes. These are given by Willey \cite{32}
as radial integrals over $j_{L}(qr)$ and a unit normalized 1p radial
density : \begin{eqnarray}
\langle j_{0}\rangle_{1p,1p}=(1-2x/3)e^{-}{}^{x}, & \langle j_{2}\rangle_{1p,1p}=(2x/3)e^{-}{}^{x}\label{eq:18}\end{eqnarray}
 in which $x=q^{2}b^{2}/4$ and $b$ is the usual three dimensional
harmonic oscillator length. In the case of $^{15}\text{N}$ there
is elastic electron scattering data available \cite{33} for the $M1$
form factor which we denote as $F_{T}(q)$. In the simple shell model
this $M1$ form factor involves a $1p_{1/2}$ proton-hole description
and the form factor is given by: \begin{equation}
F_{T}(q)=K\mu q\{(1-2x/3)+4x/9(g_{pl}+\mu_{p})/\mu\}e^{-}{}^{x}F_{SN}F_{c.m.}\label{eq:19}\end{equation}
 in which $F_{SN}$ is the nucleon size form factor and $F_{c.m.}$
for harmonic oscillator states is a simple Gaussian so that $e^{-}{}^{x}F_{c.m.}$
is replaced by $e^{-}{}^{x}{}^{(A-1)/A}$ with $A=15$. In this work
we use a simple dipole form for $F_{SN}=(1+.054675q^{2})^{-2}$ corresponding
to an rms radius of $0.81\text{ fm}$. The observed magnetic moment
$\mu$ is $-0.2831888\text{ n.m.}$ and $K=2^{-1/2}hc/(2\pi\text{Mc}{}^{2}Z)$
has the value $0.02124\text{ fm}$ for $Z=7$. The above formula has
been checked for the shell model calculations with various $b$ values
and reproduces the results given in \cite{33} using free nucleon
$g_{pl},\mu_{p}$ values. As Singhal \textit{et al} \cite{33} point
out the simple shell model calculations overestimate the peak value
of $|F_{T}|^{2}$ by 20-30\% and underestimate it for $q_{eff}$ beyond
$2.4\text{ fm}^{-1}$ by large factors. Using our renormalized $g_{pl}^{*},\mu_{p}^{*}$
in place of the free values in \eqref{eq:19} makes the shell model
calculated $|F_{T}|^{2}$ values even larger and makes no significant
improvement in the large q behavior. Only by including configurations
from the $2p_{1/2}$ shell can the data be fitted \cite{33} for all
$q$ values. 

In the cluster model for$^{15}\text{N}$ we expect the radial distribution
of the proton-hole in the highly correlated tetrahedral alpha-like
cluster $^{16}\text{O}$ reference state to be considerably different
from the $1p$ independent particle radial wave function. Lacking
a detailed theory for the many-body interactions it is convenient
to use a simple expansion for the Fourier-Bessel densities \eqref{eq:18}
as a linear superposition of $1p$ densities with different values
of the oscillator parameter:\begin{eqnarray}
F_{c.m.}\langle j_{0}\rangle_{1p,1p}=n^{-1}\sum_{n}(1-2x(n)/3)e^{-(A-1)}{}^{x(n)/A},\: & F_{c.m}\langle j_{2}\rangle{}_{1p,1p}=n^{-1}\sum_{n}2x(n)/3e^{-(A-1)}{}^{x(n)/A}\label{eq:20}\end{eqnarray}
 in which $x(n)=q^{2}b_{n}^{2}/4$ and $n$ is limited to four. The
four values of $b_{n}$ are found by fitting the $M1$ elastic electron
scattering data for $^{15}\text{N}$ and $^{14}\text{N}$. For the
$^{14}\text{N}$ case we use the similar expressions to those used
in \cite{34} \begin{equation}
F_{elT}(q)=qn^{-1}\sum_{n}e^{-(A-1)}{}^{x(n)/A}(A_{0}+A_{1}x(n))F_{SN}\label{eq:21}\end{equation}
\begin{equation}
F_{inT}(q)=qn^{-1}\sum_{n}e^{-(A-1)}{}^{x(n)/A}(B_{0}+B_{1}x(n))F_{SN}\label{eq:22}\end{equation}
with the $A_{i},B_{i}$ amplitudes being given in terms of the structure
amplitudes in \textit{L-S} coupling as\begin{equation}
A_{0}=(2/3)^{1/2}\mu K,\: A_{1}=-(2/3)^{1/2}K\mu_{d}^{*}(W_{1}-W_{2})/3\label{eq:23}\end{equation}
\begin{equation}
B_{0}=-(3/2)^{1/2}K\{2^{-1/2}(\mu_{p}^{*}-\mu_{n}^{*})V_{1}+g_{lp}^{*}V_{3}\},\: B_{1}=(2/3)^{1/2}K(\mu_{p}^{*}-\mu_{n}^{*})(V_{1}-V_{2})/3\label{eq:24}\end{equation}
in which $\mu$ is the magnetic moment of $^{14}\text{N}$ (calculated
here to be the observed value of $-0.2831888\text{ n.m.}$). The coefficients
$W_{i}$ are defined in Genz \textit{et al }\cite{11} and in our
model they are:\begin{equation}
W_{1}=2\alpha^{2}-\gamma^{2},\: W_{2}=-(4/5)^{1/2}\alpha\gamma+(27/10)^{1/2}\beta\gamma+\gamma^{2}\label{eq:25}\end{equation}
in which the $W_{2}$ differs from theirs by the coefficient of $\gamma$$^{2}$
due to the deuteron-like hole in the same alpha-like cluster having
the orbital angular momentum ($L=2$) entirely in the coordinate connecting
the center of mass of the hole pair to the center of mass of the reference
system. Note that the quadrupole moment $Q$ of $^{14}\text{N}$ is
given here by: $Q=\langle r^{2}\rangle/5\{(16/5)^{1/2}\alpha\gamma-\beta^{2}+\gamma^{2}\}$
which differs from the expression for $Q$ in Genz \textit{et al}
only by our using unity instead of 7/10 as we also used in \eqref{eq:17-1}.
The coefficients $V_{i}$ are identical to those in Genz\textit{ et
al}; $V_{1}=-2^{1/2}(\xi_{0}\alpha-\eta_{0}\beta/3^{1/2}),$ $V_{2}=-((2/5)^{1/2}\xi_{0}\gamma+6^{-1/2}\eta_{0}\beta+(9/20)^{1/2}\eta_{0}\gamma)$
and $V_{3}=-(2/3)^{1/2}\xi_{0}\beta+(2/9)^{1/2}\eta_{0}\alpha-(5/18)^{1/2}\eta_{0}\gamma$
which when used in \eqref{eq:17-1} yield with the wave functions
in section \textbf{V} the result of 0.256 n.m. as needed for the radiative
width to be $6.7\text{ meV}.$ The value of the radiative width being
$6.7\text{ meV}$ in the model used here is independent of up to $30\%$
variations in the value of $\eta$$_{0}$ . In this model the value
of $\eta_{0}$ must satisfy $\eta_{-}>\eta_{0}>\eta_{+}$ and in general
$\eta_{0}$ lies approximately half-way between $\eta_{-}$ and $\eta_{+}$.
Since the model is close to \textit{L-S} coupling for all ground state
wave functions and all the $\eta'\text{s}$ are less than $+0.2$
to yield good descriptions of the beta-decay data to the ground state
of $^{14}\text{N}$ it follows that the cluster picture with consistent
renormalized operators $g^{*}$ is in full agreement with the requirements
of CVC. 

\begin{figure}

\includegraphics[width=3.4in]{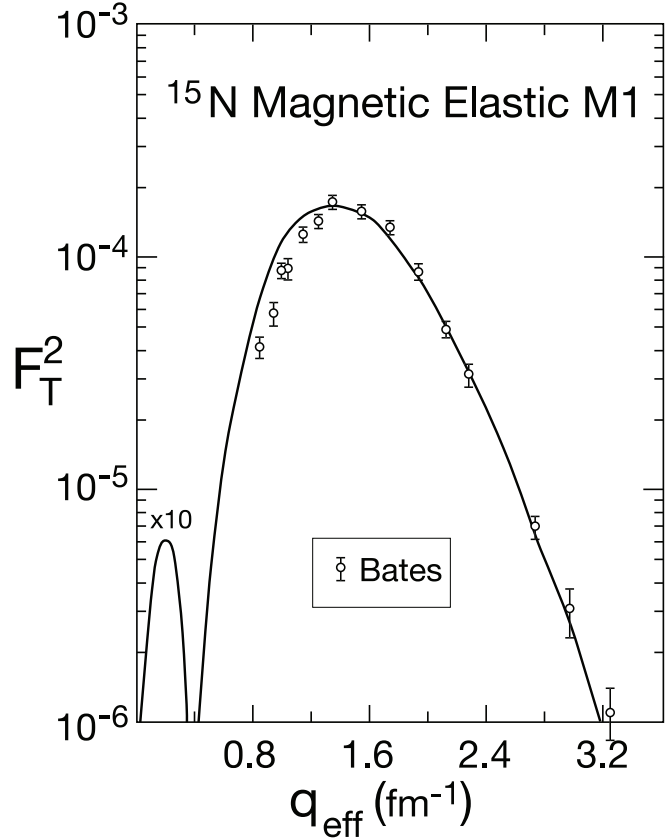}\caption{The transverse magnetic elastic form factor for $^{15}\text{N}$ calculated
using the cluster model parameters described in section \textbf{VI
}and shown by the continuous curve is compared to the Bates data \cite{33}.}
\label{fig:4}

\end{figure}

\begin{figure}

\includegraphics[width=3.4in]{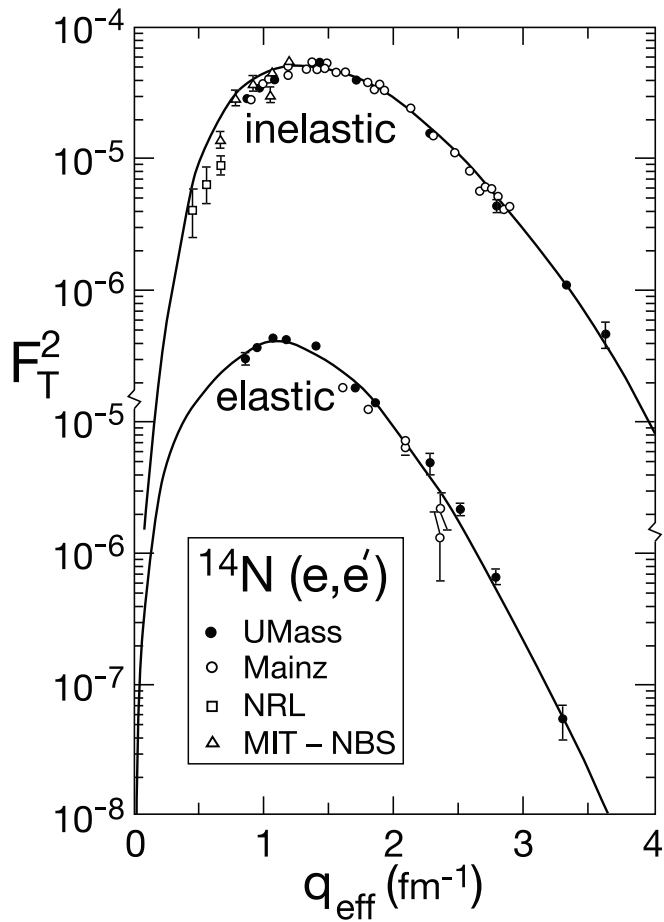}\caption{Measured elastic and inelastic $M1$ form factors for the ground state
and 2.313 MeV transition in $^{14}\text{N}$ taken from \cite{34}
(and references therein) are compared with cluster model calculations
using the parameters defined in section \textbf{VI}. }
\label{fig:5}

\end{figure}

The form factor for elastic electron $M1$ scattering has been fitted
using the set of four $b_{n}$ values (all in fm units) $b_{1}=2.85,b_{2}=1.95,b_{3}=1.82\text{ and }b_{4}=1.32$
and shows excellent agreement with the experimental data \cite{34}
in Fig. \ref{fig:4}. %
The \textit{same} set of oscillator lengths is used to calculate the
elastic electron $M1$ scattering form factor for $^{14}\text{N}$
and is compared with the experimental data in the lower panel of Fig.
\ref{fig:5}. %
 The fit is of the same quality as that shown for $^{15}\text{N}$
in Fig. \ref{fig:4}. The consistency of the model for $A=14$ and
$A=15$ gives some credence to the use of the cluster modified $1p$
density given in (20) above. The rms proton-hole radius rmsph is calculated
using $\text{rmsph}=\{5(A-1)/8(b_{1}^{2}+b_{2}^{2}+b_{3}^{2}+b_{4}^{2})/A+(0.81)^{2}\}{}^{1/2}=3.250\text{ fm }(A=15),\text{ or }3.242\text{ fm }(A=14)$
and it is the $A=14$ value we need for the calculation of $Q$ as
defined above. Using the values of $\alpha$,$\beta$, and $\gamma$
given in section \textbf{V }above we find that $Q=20.18\text{ mb}$
that agrees well with the recent experimental values of $19.3(8)\text{mb}$
and $20.01(10)\text{ mb}$ from the recent compilation \cite{35}.
A small correction to the calculated value of $Q$ arises because
of the small value of $P_{D}$ needed to fit the magnetic moment of
$^{14}\text{N}$. The estimate of this correction is uncertain but
one obtains a value of $-Q_{d}^{*}\gamma^{2}/10$ in which $Q_{d}^{*}$
is the quadrupole moment of the bound deuteron in $^{16}\text{O}$.
We anticipate that $Q_{d}^{*}$ is most likely less than the $Q_{d}$
of the free deuteron $(+2.86\text{ mb})$ so that the correction is
expected to lie between 0 and $-0.3\text{ mb}$ where the minus sign
is because of the hole nature of the deuteron. Thus our estimate of
$Q$ for $^{14}\text{N}$ is $20.03(15)$ mb where the error is comparable
to the error in the experimental value for $Q$. It is interesting
to note that the early value of $Q$ given in 1955 by Sherr \textit{et
al }was about $7\text{ mb}$ and it grew steadily over the next 38
years to the most accurate value of $20.01(10)\text{ mb}$ in 1993
. 

It is also possible to use the simple relation used in \cite{36}
between the $^{16}\text{O}$ charge radius and the $^{15}\text{N}$
charge radius for our model state for the ground state of $^{15}\text{N}$:\begin{equation}
\langle r^{2}\rangle_{ch}^{1/2}(^{15}\text{N})=\{8/7\langle r^{2}\rangle_{ch}(^{16}\text{O})-(\text{rmsph})^{2}/7\}{}^{1/2}=\{8/7(2.71)^{2}-(3.25)^{2}/7\}{}^{1/2}=2.62\text{ fm}\label{eq:26}\end{equation}
 which agrees very well with the result given for the charge radius
of $^{15}\text{N}$ in \cite{36}. 

The inelastic electron scattering from $^{14}\text{N}$ leading to
the first excited state at $E=2.313\text{ MeV}$ shows much more deviation
from the shell model calculations than the elastic data does and because
this transition involves a change of isospin from $T=0$ to $T=1$
it has the connection to beta-decay as indicated in earlier work concerned
with CVC. As we noted earlier for the beta-decay $C(Z,W)$ shapes
we need a knowledge of the average oscillator length squared which
we denote by $b_{c}^{2}$ and which we can assume should be the average
of a set of $b_{n}^{2}$ that fit the inelastic electron scattering
form factor. Our fit to the data using (all in fm units) $b_{1}=3.512,b_{2}=2.12,b_{3}=1.50,\text{ and }b_{4}=1.09$
describes the data very well including the large q values. The value
of $b_{c}$ from $b_{c}^{2}=(b_{1}^{2}+b_{2}^{2}+b_{3}^{2}+b_{4}^{2})/4$
is 2.25 fm which was used to calculate the shapes for the beta decays
discussed in section \textbf{V }above.

\section{GAMMA DECAYS IN $^{14}\text{N}$}

In most of the work on beta decay in the $A=14$ system only the $M1$
decay of the first excited state has been calculated. In this section
we focus on the gamma decay of the $J^{P}=1^{+}$ at $E=3.948\text{ MeV}$
excitation. The gamma decay to the ground state involves two multipoles
corresponding to $E2$ and $M1$. The wave functions in section \textbf{V}
gives the values of 0.0026 eV and 0.00042 eV for the radiative widths
for $E2$ and $M1$ respectively which agree well with the corresponding
experimental values 0.003 eV and 0.0004 eV from the TUNL compilation
\cite{37}. The gamma decay of the 3.948 MeV state to the $T=1,0^{+}$
state (2.313 MeV) is the almost completely dominant decay mode and
is pure M1. The wave functions in section V give a radiative width
of 0.155 eV which is in good agreement with the experimental value
of 0.091(30) eV. Overall the cluster model calculated gamma decay
widths of the 3.948 MeV state are in satisfactory agreement with experiment
but much of the gamma decay data in $^{14}\text{N}$ is quite old
and new measurements could provide more accuracy on the transitions
between the low-lying positive parity states.

\section{DISCUSSION AND CONCLUSIONS}

The results given in the preceding sections represent a convincing
argument that the standard simple shell model is not an optimal
starting point for describing the ground state of  $^{16}\text{O}$
which, in turn, means it is not an optimal starting point for describing
the low-lying states of $A=15$ and $A=14$ nuclei. In particular
the cluster model need for the $0^{+}$ ground states of $A=14$ nuclei
with $T=1$ to be more than $97\%$ in the $^{1}S_{0}$ configuration
(corresponding in \textit{j-j} coupling to 1/3 probability for the
$p_{1/2}^{-2}$ component and 2/3 probability for the $p_{3/2}^{-2}$
component ) contradicts the expectations of the strong spin-orbit
j-j shell model ideology as pointed out by Talmi \cite{8}. More important
however is that the strong correlations leading to alpha-like clustering
with $T_{d}$ point group symmetry is that $^{16}\text{O}$ is a tetrahedrally
deformed nucleus and not spherical, as all shell model calculations
use as a basic starting point. It is strongly deformed as the ground
state rotational band \cite{19} with the sequence $0^{+},3^{-},4^{+},6^{+},$
has strongly enhanced $BE3,BE4$ values for the transitions from the
$3^{-},4^{+}$ states respectively to the ground state that are typical
of a simple tetrahedral rotor model. There is no quadrupole deformation
in the tetrahedral model of the ground state rotational band of $^{16}$
O as it violates the boson symmetry for four identical alpha-like
clusters (tetrons?). \textit{We believe for $N=Z$ even-even nuclei
that the clustering dynamics of $(N+Z)/4$ identical tetron clusters
determines the various multipole deformations with $L>1$ for each
nucleus.} \textit{ }

We note that the next alpha-like cluster system to not have any quadrupole
deformation in the lowest energy intrinsic state is $^{40}\text{Ca}$
as it appears to also be tetrahedrally deformed rather than spherical.
In particular the lowest lying excited states are $0^{+}(E=3.35\text{ MeV})$
and $3^{-}(E=3.74\text{ MeV})$ with the $BE3$ value being 31 (W.U).
Surprisingly at first is the fact that the binding energy of the last
neutron (15.643 MeV) in $^{40}\text{Ca}$ is almost identical to the
binding energy (15.664 MeV) of the last neutron in $^{16}\text{O}$
. The difference is 21 keV and we do not for one moment believe this
is an accidental degeneracy. We conjecture that the last neutron taken
from one of the outermost tetron clusters will have three neighboring
tetron clusters which are in the same close packed configuration as
the four tetrons in $^{16}\text{O}$. The remaining six tetrons in
a tetrahedral $^{40}\text{Ca}$ are all spatially removed from the
four containing the last neutron so that there is almost no interaction
between the six spectators and the neutron being taken out. In short
the similarity in the neutron binding energy in the tetrahedral model
arises because the weakest bound neutron interacts only with the nearest
neighbor clusters which is the same in $^{16}\text{O}$ and $^{40}\text{Ca}$.
Of course this does not happen for the binding energy of the last
proton because it sees the long range Coulomb interaction from all
the spectators. 

In summary the new solution to the $A=14$ beta-decay puzzle uses a 
highly correlated model in which the unperturbed LSJT basis has a
simple orbital angular momentum selection rule which forbids the 
ground state GT transitions in the mirror $\beta$-decays. This rule arises
in this model because the $L=2$, $S=1,$ $J=1,$ $T=0$ state is the most 
appropriate assignment for $^{14}$N$_\text{gs}$ and the initial mirror $T=1$ ground
states in the p$^{-2}$ basis space (even with mixing) has no $L=2$ component.     
Only by mixing in $L=1$ and $L=0$ states in the $T=0$ sector can the
mirror ground states have non-zero MGT elements. In the shell model 
the use of reasonably strong tensor and spin-orbit interactions causes 
the $L=2$ state in $^{14}$N$_\text{gs}$ to become significantly mixed with the $L=0$ and 
$L=1$ basis states which has to be fine tuned to cancel the MGT elements when the $T=1,$ $L=0$ and $L=1$ states in $^{14}$C and $^{14}$O are automatically 
strongly mixed by the strong spin-orbit interaction. The nature of the reference $^{16}$O state in this alternative model leads to the expectation of 
a strong suppression of the nuclear spin-dependent mixing interactions which is why the model has weak mixing between the LSJ states for 
both $T=0$ and $T=1$ states and changes the infinite life of $^{14}$C into a long
lifetime. The model explains the large asymmetry between the mirror
$\beta$-decays because of the interference between the nuclear and electromagnetic spin-orbit mixing interactions with the latter term having
opposite sign for the state in $^{14}$O to that in $^{14}$C. The MGT elements
for these two transitions are indeed very different in magnitude with 
this model and also their MGT elements have opposite sign in order
to obtain the observed negative values for the $a_+$, $a_-$ shape slope parameters. Hopefully the detailed results provided in this work on the $\beta$-decays
in $A=14$ nuclei will be helpful to future investigators in their search for
a more complete description of the structure of p-shell nuclei. 
 
\acknowledgements{ My
colleagues at Florida State University, in particular K. Kemper, A.
Volya, V. Abramkina, and S. Tabor are greatly acknowledged. Special thanks for several
discussions are due to J. C. Hardy, B. A. Brown, A. Richter and D.
J. Millener. }

\end{document}